\documentclass[12pt]{iopart}
\usepackage{graphicx}
\usepackage{dcolumn}
\usepackage{bm}
\usepackage{epsf}

\newcommand{\xiv}{\mbox{\boldmath$\xi$}}
\newcommand{\sigv}{\mbox{\boldmath$\sigma$}}

\newcommand{\alphav}{\mbox{\boldmath$\alpha$}}
\newcommand{\piv}{\mbox{\boldmath$\pi$}}

\begin{document}

\title{Including Nuclear Degrees of Freedom in a Lattice Hamiltonian}

\author{Peter L. Hagelstein$^1$ and Irfan U. Chaudhary$^2$}

\address{$^1$Research Laboratory of Electronics, \\ 
Massachusetts Institute of Technology,       \\
Cambridge, MA 02139, USA                     \\
E-mail: plh@mit.edu}

\address{$^2$Department of Computer Science and Engineering, \\
             University of Engineering and Technology. \\
             Lahore, Pakistan}

\begin{abstract}

Motivated by many observations of anomalies in
condensed matter systems, we consider a new fundamental
Hamiltonian in which condensed matter and nuclear systems are
described initially on the same footing. Since it may be possible
that the lattice will respond to the mass change associated with a
excited nuclear state, we adopt a relativistic description
throughout based on a many-particle Dirac formalism.  This
approach has not been used in the past, perhaps due to the
difficulty in separating the center of mass and relative degrees
of freedom of the nuclear system, or perhaps due to an absence of
applications for such a model.  
In response to some recent ideas about how to think about the
center of mass and relative separation, we obtained from the
Dirac model a new fundamental Hamiltonian in which the lattice
couples to different states within the composite nuclei 
within the lattice.
In this description the different nuclear states have 
different mass energies and kinetic energies, as we had
expected.
In addition there appear new terms which provide for nuclear
excitation as a result of coupling to the composite momentum.
This new effect comes about because of changes in the 
composite nuclear state as a result of the dynamical Lorentz boost
in the lattice.

\end{abstract}

\section{Introduction}

From a condensed matter viewpoint, a solid is made of nuclei and
electrons, where the nuclei can for the most part be treated as
point particles  \cite{Kittel}.  In cases where a more sophisticated
description is needed, the basic description is augmented with
nuclear spin, magnetic moments, and electric quadrupole moments
\cite{Abragam}. Essentially the only place that excited states show up is in
M\"ossbauer studies where they are required in order to describe
the absorption or emission of a gamma \cite{Frauenfelder}.
The picture that results is both wonderfully simple and very rich;
simple, in that the Born-Oppenheimer separation allows us to reduce the problem
to electronic bands and phonon modes;
and rich, as the models that result describe a wide range of basic, subtle, and
occasionally unexpected physical phenomena.
This basic approach to condensed matter physics has been sufficiently
successful over the years that it would require some rather dramatic
new experimental result before we might be motivated to revise it in
any significant way.

In recent years there have been claims of experimental results
which at a fundamental level seem not to be consistent with this
basic viewpoint of condensed matter physics.  In the
Fleischmann--Pons excess heat effect \cite{Flei89,Flei90}, a great deal of energy
is generated which is thought to have a nuclear origin (since
there are no commensurate chemical byproducts, and since $^4$He is
observed as a possible product in amounts proportional to the
energy produced \cite{Miles93,Miles94,Miles04}), without commensurate energetic nuclear
products \cite{Hagelstein2010}. In these experiments it almost seems as if the solid
is taking up an MeV quantum; if so, then this constitutes an
effect which seems very hard to understand within our current
condensed matter framework. Given that such an effect seems
impossible to contemplate within modern condensed matter physics
(and also within modern nuclear physics), a natural reaction has
been to go with the existing picture (supported by a very large
body of experimental results and a consistent and mature theory),
and to reject the Fleischmann--Pons experiment as simply being in
error \cite{Fiasco}.

During the past two decades and more this de facto solution has
been adopted generally, and it has worked surprisingly well.
Science has advanced substantially; there are now even more
experimental and theoretical results which support the modern
condensed matter viewpoint; and those who pursue anomalies such as
the Fleischmann--Pons experiment are isolated from science and
ignored \cite{Demarcating}.

Meanwhile, another experimental result has been put forth which
challenges our modern view of condensed matter physics.  Karabut
has studied a variety of anomalies in high-current density glow
discharge experiments, and in the course of the work noticed that
collimated X-ray emission occurred in powerful bursts normal to
metal coin-shaped samples that served as cathodes \cite{Karabut2002,Karabut2003,Karabut2004,Karabut2005}.
Although anomalous emission effects are seen when the discharge is
on, the powerful bursts of collimated X-rays are observed on the
order of a millisecond after the discharge is turned off.  A
current spike occurs when the discharge is shut off, which we
might imagine causes vibrations in the sample.  It seems as if the
vibrational energy is being communicated somehow to produce
in-phase electronic or nuclear excitation at X-ray energies, which
then produces collimated X-ray emission through a phased array
emission effect.  We note that a related effect involving the
collimated emission of gamma-rays in beamlets was reported earlier
by Gozzi \cite{Gozzi1998}. Needless to say, such an effect has no place in
modern condensed matter physics.

In experiments performed by the Piantelli group, hydrogen is
absorbed in nickel samples at elevated temperature, resulting in a
thermal effect (consistent with energy generation) \cite{Focardi1994},\cite{Focardi1998},
low-level nuclear effects (gamma and neutron emission  \cite{Battaglia1999},\cite{Campari2000}),
and the appearance of new elements \cite{Campari2004}. This latter effect
(appearance of new elements) in these experiments is not a
low-level effect. Once again, such effects are not predicted in
modern condensed matter physics.

These experimental results, and many others, have motivated us to explore new models that might be
relevant.
A major issue that we have been interested in is the possibility of coherent energy exchange between quantum systems
with mismatched characteristic energies, which we considered to be the biggest theoretical problem
associated with the anomalies.
Coherent energy exchange between mismatched quantum systems occurs
in high harmonic generation \cite{Dromey2007}, so we know that it is possible
in principle.
However, there seems to be no analog to Corkum's mechanism \cite{Corkum1993},\cite{Lewenstein1994}
present in the condensed matter system.
A lesser version of the effect is known within the multiphoton
regime of the spin-boson model, which is used to model basic
linear interactions of two-level systems with an oscillator
\cite{Bloch1940},\cite{Shirley1965},\cite{Cohen1973}.  We found that if the two-level system is augmented with
loss, the coherent energy exchange rate is increased dramatically.
This is due to the fact that destructive interference limits the
rate at which coherent energy exchange occurs in the spin-boson
model, so augmenting the model with a mechanism that removes this
destructive interference would be expected to improve coherent
energy exchange rates \cite{CMNS5},\cite{CMNS7},\cite{CMNS9},\cite{CMNS10}.

Coherent energy exchange in these models works best when the
coupling between the two-level transition (representing electronic
and nuclear transitions) and oscillator (representing a
vibrational mode) is strong.  We studied a further generalization
of the lossy spin--boson model in which two transitions are
coupled to an oscillator, where one is strongly coupled and one is
weakly coupled \cite{CMNS11}.  We found that the strongly coupled system
could assist coherent energy exchange for the weakly coupled
system.  The model that resulted appeared to us to be very closely
related to excess heat production in the Fleischmann--Pons
experiment, assuming that the mechanism involved D$_2$/$^4$He
transitions that were weakly coupled to a phonon mode (weakly
coupled due to the Coulomb repulsion between the deuterons), and
that a strongly coupled transition were also present.  The big
problem in this kind of model ends up being the identification of
the strongly coupled transition.  Finding an appropriate strongly
coupled transition with sufficiently strong coupling to do the job
seems problematic within the approach \cite{Birdseye}.

After analyzing many candidate transitions, we came to the
conclusion that there were no physical transitions which could
serve as the strongly coupled two-level transition within the
model.  We were optimistic in our writing about the possibility
that systems described by three-level systems (or $N$-level
systems) would be able to do the job.  After putting in a great
deal of work on analyzing the strongly coupled three-level system,
it seemed once again that the coupling was simply not strong
enough to make a connection with experiment.  This conclusion was
supported by spectral data from Karabut \cite{Karabut2012}, which seemed to be
qualitatively consistent with the approach and models, but which
would require much stronger coupling to explain.

All of this has led us to the conclusion that we are going to need a new kind of model in order to
account for the experimental results.  To obtain coupling sufficiently strong to be consistent with
the Karabut experiment, we require very strong interactions that are on the general order of what
would occur in a nuclear configuration mixing calculation.  Yet there is no reason to expect that
nuclear configuration interactions can couple to a phonon mode.  In our earlier efforts to describe
such an effect, we concluded that the internal nuclear degrees of freedom associated with configuration
mixing separated cleanly from the vibrational degrees of freedom.  For many years there has seemed to
be no viable solution within the general approach, which has been very discouraging.

The intuitive picture that has emerged over the past few years of thinking about the problem is
that the different excited states of the nucleus have different masses, and under appropriate
conditions it may be possible for the nucleus to notice the mass differences of the different
configurations.  This could provide the physical basis for phonon exchange in association with
configuration mixing.  To describe such an effect, we need to develop a description of the
associated coupling, which seems not to be available in the literature.  One approach is to
begin with a relativistic model for the nucleons, and then reduce it in some way to obtain
a low momentum approximation in which the associated mass effects are retained.

The issues under discussion fit within the generic heading of
relativistic quantum mechanics, which before 1950 would likely
have implied the two-body Dirac equation as a starting place.
However, the need for a manifestly covariant relativistic quantum
theory more generally led to the development of modern quantum
field theory, which could in principle be used for problems of
interest to us.  Field theory is much more complicated than
relativistic quantum mechanics, so we would prefer a simpler model
derived from relativistic quantum mechanics if possible. In this
day and age, there are many relativistic quantum models that have
been derived from field theory (such as the Bethe--Salpeter
equation, as well as others \cite{Gross1999}).

The separation of relative and center of mass degrees of freedom
that is possible in the nonrelativistic problem does not extend to
the relativistic problem, which complicates things.  In a recent
manuscript \cite{Kinematic} we obtained a result with the two-body Dirac model
that seemed to suggest it was possible to arrange for a separation
of the relative and center of mass contributions to the energy in 
a simple way.  We originally made use of this approach in the 
this paper to obtain an approximate model for the center of mass
and relative dynamics discussed in this paper.  However, in 
responding to the reviewers' comments on the paper, it became clear 
that there were weaknesses in the approach outline, and our
interpretation of the results have changed.  We were able to 
revise this paper in the galley proof stage in order to take into 
account the improved point of view.  The model that results (which is
essentially the same in both cases) can then be used
directly to develop a new Hamiltonian for nuclei in a lattice that
includes the coupling consistent with a many-particle Dirac
formulation.

Interestingly, the model that results seems to include a relativistic effect which provides
a direct coupling between the lattice motion and excitations in the nucleus.   The resulting
model appears to be much more closely connected with our earlier models than we had
expected, which provides the motivation to explore the model further in the future.


\section{Center of mass and relative contributions to the energy}

We begin with a description of the free nucleus in terms of Dirac particles within the
context of a many-particle Dirac Hamiltonian

\begin{equation}
\hat{H} ~=~ \sum_j \alphav_j \cdot c \hat{\bf p}_j + \beta_j M_j
c^2 + \sum_{j<k} V_{jk}({\bf r}_k - {\bf r}_j).
\end{equation}

\noindent
Protons and neutrons are composite particles with internal quark structure, and
one might criticize the use of a Dirac point-particle description for composites in
this case; however, for our purposes it seems the simplest place to start.  The
Dirac $\alphav$ and $\beta$ matrices are

\begin{equation}
\alphav_j
~=~
\left (
\begin{array} {cc}
0 & \sigv_j \cr
\sigv_j & 0 \cr
\end{array}
\right )_j, \ \ \ \ \ \ \ \ \ \ \beta ~=~ \left (
\begin{array} {cc}
I & 0 \cr
0 & -I \cr
\end{array}
\right )_j.
\end{equation}

\noindent
The interaction between two nucleons appears here as $V_{jk}({\bf r}_k - {\bf r}_j)$;
we assume that this includes strong force and electromagnetic interactions.
We assume the $\Phi$ is an exact solution to the time-independent
equation

\begin{equation}
E \Phi ~=~ \bigg [ \sum_j \alphav_j \cdot c \hat{\bf
p}_j + \beta_j M_j c^2 + \sum_{j<k} V_{jk}({\bf r}_k - {\bf r}_j)
\bigg ] \Phi. \label{eig1}
\end{equation}

\subsection{Center of mass and relative coordinates}

The classical center of mass coordinate satisfy

\begin{equation}
M {\bf R} ~=~ \sum_j m_j {\bf r}_j
\end{equation}

\noindent
with

\begin{equation}
M  ~=~ \sum_j m_j.
\end{equation}

\noindent
The relative position coordinates are

\begin{equation}
\xiv_j ~=~ {\bf r}_j - {\bf R}.
\end{equation}

\noindent
The total classical momentum is

\begin{equation}
{\bf P} ~=~ \sum_j {\bf p}_j
\end{equation}

\noindent
and the relative momenta are

\begin{equation}
\piv_j ~=~ {\bf p}_j - {m_j \over M} {\bf P}.
\end{equation}

\noindent
One of the relative position operators is redundant, since the sum of all relative position operators is zero;
similarly one of the relative momentum operators is redundant.

\subsection{Mass parameter}

The eigenvalue can be expressed in terms of relative and center of mass matrix elements according to

$$
E
~=~
\bigg \langle \Phi \bigg |
\left ( \sum_j {m_j \over M} \alphav_j \right ) \cdot c   \hat{\bf P}
\bigg | \Phi \bigg \rangle
\ \ \ \ \ \ \ \ \ \ \ \ \ \ \ \ \ \ \ \ \ \ \ \ \ \ \ \ \ \
\ \ \ \ \ \ \ \ \ \ \ \ \ \ \ \ \ \ \ \ \ \ \ \ \ \ \ \ \ \
$$
\begin{equation}
+ \bigg \langle \Phi \bigg | \sum_j \alphav_j \cdot c \hat{\piv}_j
+ \sum_j \beta_j m_jc^2 + \sum_{j<k} V_{jk}(\xiv_k-\xiv_j) \bigg |
\Phi \bigg \rangle,
\end{equation}

\noindent where $\Phi$ is an exact solution to the eigenvalue
equation (Eq. (3)). We add and subtract mass terms to obtain

$$
E
~=~
\bigg \langle \Phi \bigg |
\left ( \sum_j {m_j \over M} \alphav_j \right ) \cdot c   \hat{\bf P}
+
\left ( \sum_j {m_j \over M} \beta_j \right ) M^* c^2
\bigg | \Phi \bigg \rangle
\ \ \ \ \ \ \ \ \ \ \ \ \ \ \ \ \ \ \ \ \ \ \ \ \ \ \ \ \ \
\ \ \ \ \ \ \ \ \ \ 
$$
\begin{equation}
+ \bigg \langle \Phi \bigg | \sum_j \alphav_j \cdot c \hat{\piv}_j
+ \sum_j \beta_j m_jc^2 + \sum_{j<k} V_{jk}(\xiv_k-\xiv_j) - \left
( \sum_j {m_j \over M} \beta_j \right ) M^* c^2 \bigg | \Phi \bigg
\rangle.
\end{equation}

\noindent
If we define the mass parameter $M^*$ according to

\begin{equation}
M^*c^2 ~=~ { \displaystyle{ \bigg \langle \Phi \bigg | \sum_j
\alphav_j \cdot c \hat{\piv}_j + \sum_j \beta_j m_jc^2 +
\sum_{j<k} V_{jk}(\xiv_k-\xiv_j) \bigg | \Phi \bigg \rangle }
\over \displaystyle{ \bigg \langle \Phi \bigg | \sum_j {m_j \over
M} \beta_j \bigg | \Phi \bigg \rangle } },
\end{equation}

\noindent
then the associated matrix element for the modified relative contribution vanishes.
The idea here is that the contribution of the relative problem now appears in the 
mass parameter associated with the center of mass contribution to the eigenvalue.

\subsection{Eigenvalue relation for a single configuration}

With this definition of the mass  parameter the eigenvalue simplifies to

\begin{equation}
E ~=~ \bigg \langle \Phi \bigg | \left ( \sum_j {m_j
\over M} \alphav_j \right ) \cdot c   \hat{\bf P} + \left ( \sum_j
{m_j \over M} \beta_j \right ) M^* c^2 \bigg | \Phi \bigg \rangle.
\end{equation}

\noindent
We can rotate to obtain

\begin{equation}
E ~=~ \bigg \langle \Phi' \bigg | \sum_j {m_j \over M}
\beta_j \bigg | \Phi' \bigg \rangle \sqrt{(M^* c^2)^2 + c^2 |{\bf
P}|^2} \label{relation},
\end{equation}

\noindent
where we used

\begin{equation}
\Phi' ~\sim~ \textrm{e}^{i {\bf P} \cdot {\bf R} / \hbar}.
\end{equation}

\noindent These are the basic results that we reported in \cite{Kinematic}. However, our interpretation
now is different.  The mass parameter $M^*$ can allow us to connect the center of mass
problem for the interacting Dirac particles with the center of mass problem of
a set of equivalent non-interacting Dirac particles that have a different rest mass.

\subsection{Composite masses}

In general the mass parameter $M^*$ for the composite will in general depend on
${\bf P}$ since the Dirac Hamiltonian is in general not covariant (it has not yet
been shown that $M^*$ should be invariant for a covariant model, but we think it
likely).  Also, in a covarient model we would expect the energy-momentum relation to
be

\begin{equation}
E ~=~ \sqrt{ (M_0c^2) + c^2 |{\bf P}|^2}
\end{equation}

\noindent
where $M_0$ is the composite rest mass.
In light of this, we could correct the non-covariant Dirac model by requiring

\begin{equation}
\bigg \langle \Phi' \bigg | 
\sum_j {m_j \over M} \beta_j 
\bigg | \Phi' \bigg \rangle 
\sqrt{(M^* c^2)^2 + c^2 |{\bf P}|^2}
~\to~
\sqrt{ (M_0c^2) + c^2 |{\bf P}|^2}
\end{equation}

\noindent
In this case, we can recover the nonrelativistic limit

\begin{equation}
E ~\to~ M_0 c^2 + {|{\bf P}|^2 \over 2 M_0} + \cdots
\end{equation}


\section{Finite Basis State Model}

We would like to expand our description to include a finite set of states, and in the process
we would like a formulation in which the state mass impacts the kinematics.  We begin by
assuming a finite basis state model of the form

\begin{equation}
\Psi ~=~ \sum_j c_j \Phi_j.
\end{equation}

\noindent
We make use of the variational method to obtain the matrix equation

\begin{equation}
E{\bf c} ~=~ {\bf H} \cdot {\bf c},
\end{equation}

\noindent
where the matrix ${\bf H}$ has individual matrix elements given by

\begin{equation}
H_{kl} ~=~ \langle \Phi_k | \hat{H} | \Phi_l \rangle.
\end{equation}

\noindent
The vector ${\bf c}$ made up of the different expansion coefficients is

\begin{equation}
{\bf c}
~=~
\left (
\begin{array} {c}
c_1 \cr
c_2 \cr
c_3 \cr
\vdots \cr
\end{array}
\right ).
\end{equation}

\noindent
This coefficient vector ${\bf c}$ should not be confused with the speed of light $c$.

\subsection{Diagonal matrix elements and mass parameters}

To arrange for dynamics with the mass parameter matched to the
state, we focus first on the diagonal matrix elements and write

\begin{equation}
H_{kk} ~=~ \langle \Phi_k | \hat{H}_{\bf R} | \Phi_k \rangle +
\langle \Phi_k | \hat{H}_{\bf r} | \Phi_k \rangle,
\end{equation}

\noindent
where the center of mass and relative parts are

\begin{equation}
\langle \Phi_k | \hat{H}_{\bf R} | \Phi_k \rangle ~=~ \bigg
\langle \Phi_i \bigg | \left ( \sum_j {m_j \over M} \alphav_j
\right ) \cdot c   \hat{\bf P} + \left ( \sum_j {m_j \over M}
\beta_j \right ) M^*_k c^2 \bigg | \Phi_k \bigg \rangle,
\end{equation}

\begin{equation}
\langle \Phi_k | \hat{H}_{\bf r} | \Phi_k \rangle ~=~ \bigg
\langle \Phi_k \bigg | \sum_j \alphav_j \cdot c \hat{\piv}_j +
\sum_j \beta_j m_jc^2 + \sum_{j<k} V_{jk}(\xiv_k-\xiv_j) - \left
(\sum_j {m_j \over M} \beta_j \right )M^*_k c^2 \bigg | \Phi_k
\bigg \rangle.
\end{equation}

\noindent
If we require the relative contribution to vanish (as discussed above), then the state-dependent
mass parameter  $M_k^*$ is consistent with

\begin{equation}
M_k^* c^2 ~=~ { \displaystyle{ \bigg \langle \Phi_k \bigg | \sum_j
\alphav_j \cdot c \hat{\piv}_j + \sum_j \beta_j m_jc^2 +
\sum_{j<k} V_{jk}(\xiv_k-\xiv_j) \bigg | \Phi_k \bigg \rangle }
\over \displaystyle{ \bigg \langle \Phi_k \bigg | \sum_j {m_j
\over M} \beta_j \bigg | \Phi_k \bigg \rangle } }.
\end{equation}

\subsection{Off-diagonal matrix elements}

The off-diagonal matrix elements may be written as

$$
H_{kl}
~=~
\bigg \langle \Phi_k \bigg |
 \sum_j {m_j \over M} \alphav_j
\cdot c   \hat{\bf P}
\bigg | \Phi_l \bigg \rangle
\ \ \ \ \ \ \ \ \ \ \ \ \ \ \ \ \ \ \ \ \
\ \ \ \ \ \ \ \ \ \ \ \ \ \ \ \ \ \ \ \ \
\ \ \ \ \ \ \ \ \ \ \ \ \ \ \ \ \ \ \ \ \
$$
\begin{equation}
+ \bigg \langle \Phi_k \bigg | \sum_j \alphav_j \cdot c
\hat{\piv}_j + \sum_j \beta_j m_jc^2 + \sum_{j<k}
V_{jk}(\xiv_k-\xiv_j) \bigg | \Phi_l \bigg \rangle,
\end{equation}

\noindent
where the first term is associated with the center of mass, and the
second term is associated with the relative problem.

\subsection{Finite basis eigenvalue relations}

The finite basis eigenvalue relations for a composite in free space
can be written in the form

\begin{equation}
E c_k ~=~ \left [ \bigg \langle \Phi_k' \bigg |
\sum_j {m_j \over M} \beta_j \bigg | \Phi_k' \bigg \rangle \sqrt{
(M_k^*c^2)^2 + c^2 |{\bf P}|^2} \right ] c_k + \sum_{l \ne k}
H_{kl} c_l.
\end{equation}

\noindent
It will be convenient to write the off-diagonal matrix element in this case as

\begin{equation}
H_{kl}
~=~
\overline{\alphav}_{kl} \cdot
 ( c  {\bf P} )
+ V_{kl},
\end{equation}

\noindent
where $V_{kl}$ is the coupling matrix element from the relative part of the problem

\begin{equation}
V_{kl}
~=~
\bigg \langle \Phi_k \bigg |
\sum_j \alphav_j \cdot c \hat{\piv}_j
+
\sum_j \beta_j m_jc^2
+
\sum_{j<k} V_{jk}(\xiv_k-\xiv_j)
\bigg | \Phi_l \bigg \rangle
\end{equation}

\noindent
and where the vectors $\overline{\alphav}_{kl}$ are defined by

\begin{equation}
\overline{\alphav}_{kl}
~=~
\bigg \langle \Phi_k \bigg |
 \sum_j {m_j \over M} \alphav_j
\bigg | \Phi_l \bigg \rangle.
\end{equation}

\noindent
We consider in the Appendix the nonrelativistic reduction of a representative
$\overline{\alphav}_{kl} \cdot c \hat{\bf P}$ matrix element.

\subsection{Rest frame eigenvalue equations}

The eigenvalue equations above are unfamiliar and moderately complicated. It
is useful to consider them in the rest frame; in this case, we obtain

\begin{equation}
E c_k ~=~ \left [ \bigg \langle \Phi_k \bigg | \sum_j
{m_j \over M} \beta_j \bigg | \Phi_k \bigg \rangle M_k^*c^2 \right
] c_k + \sum_{l \ne k} V_{kl} c_l.
\end{equation}

\noindent
For such a finite basis state model we would expect the basis state masses to
appear; consequently we may write

\begin{equation}
M_k c^2  ~=~ \bigg \langle \Phi_k
\bigg | \sum_j {m_j \over M} \beta_j \bigg | \Phi_k \bigg \rangle
M_k^*c^2.
\end{equation}

\noindent
This allows us to write

\begin{equation}
E
\left (
\begin{array} {c}
c_1 \cr
c_2 \cr
c_3 \cr
\vdots \cr
\end{array}
\right )
~=~
\left (
\begin{array} {cccc}
M_1 c^2 & V_{12} & V_{13} & \cdots  \cr 
V_{21} & M_2 c^2 & V_{23} & \cdots  \cr 
V_{31} & V_{32} & M_3c^2 & \cdots  \cr 
\vdots & \vdots  & \vdots &
\vdots \cr
\end{array}
\right )
\cdot
\left (
\begin{array} {c}
c_1 \cr
c_2 \cr
c_3 \cr
\vdots \cr
\end{array}
\right ).
\end{equation}

\noindent
It will be convenient to think of this as the basic unperturbed problem

\begin{equation}
E{\bf c} ~=~ {\bf H}_0 \cdot {\bf c}.
\end{equation}

\subsection{Low-momentum eigenvalue relations}

We can expand the square root terms to lowest order to obtain an eigenvalue equation
relevant to the low-momentum case; we write

\begin{equation}
E c_k
~=~
\bigg [
M_kc^2
+
{|{\bf P}|^2 \over 2 M_k}
+
\cdots
\bigg ]
c_k
+
\sum_{l \ne k}
\bigg [ \overline{\alphav}_{kl} \cdot (c {\bf P}) + V_{kl} \bigg ]
 c_l,
\end{equation}

\noindent
where we have approximated the Dirac model as covariant in the estimates
of the diagonal matrix elements.
In matrix notation, this might be written as

\begin{equation}
E {\bf c} ~=~ \bigg [ {\bf M}c^2 + {\bf
M}^{-1}  { |{\bf P}|^2 \over 2} + {\bf a} \cdot (c {\bf
P}) \bigg ] \cdot {\bf c},
\end{equation}

\noindent
where

\begin{equation}
{\bf M} ~=~ \left (
\begin{array} {cccc}
M_1 & 0 & 0 & \cdots  \cr 0 &
M_2 & 0 & \cdots  \cr 0 & 0 &
M_3 & \cdots  \cr \vdots & \vdots  & \vdots &
\vdots        \cr
\end{array}
\right )
\end{equation}

\noindent
and where

\begin{equation}
{\bf a}
~=~
\left (
\begin{array} {cccc}
0 & \overline{\alphav}_{12} & \overline{\alphav}_{13} & \cdots  \cr
\overline{\alphav}_{21} & 0 & \overline{\alphav}_{23} & \cdots  \cr
\overline{\alphav}_{31} & \overline{\alphav}_{32} & 0 & \cdots  \cr
\vdots & \vdots  & \vdots & \vdots        \cr
\end{array}
\right ).
\end{equation}


\section{A Hamiltonian for Nuclei in a Lattice}

There has been discussion over the years as to develop a suitable
formalism that would be capable of systematically addressing the
anomalies of interest in condensed matter nuclear science. It was
proposed in \cite{Formulation} that one begin with a fundamental Hamiltonian
based on nucleons and electrons, and then reduce the model for
applications by first building up nuclei from nucleons, then
solving for electronic degrees of freedom in a Born--Oppenheimer
picture, and finally focusing on the vibrational problem.  We
imagine a similar approach here, only instead of starting with a
nonrelativistic fundamental Hamiltonian we use a relativistic one.

The separation of center of mass and relative degrees of freedom
is straightforward in the nonrelativistic version of the problem,
which is why we focused on it in \cite{Formulation}.  But we see in the
discussion above that it is possible to separate the center of
mass and relative Hamiltonians for a many-particle Dirac model,
even in the context of a finite basis approximation.  This
separation allows us to extend the earlier program systematically
to a relativistic formulation (including now a new relativistic
coupling between the nuclear motion and internal nuclear degrees
of freedom) based on an underlying many-particle Dirac model.

\subsection{Hamiltonian for electrons and nucleons}

We begin with a formal model based on many-particle Dirac Hamiltonians
for the electrons and nucleons

$$
\hat{H} ~=~ \bigg [ \sum_j \alphav_j \cdot c \hat{\bf p}_j +
\beta_j M_j c^2 + \sum_{j<k} V_{jk}^{nn}({\bf r}_k - {\bf r}_j)
\bigg ]_{\textrm{nucleons}} \ \ \ \ \ \ \ \ \ \ \ \ \ \ \ \ \ \ \
\ \ \ \ \ \ \ \ \ \ \ \ \ \
$$
\begin{equation}
+ \bigg [ \sum_j \alphav_j \cdot c \hat{\bf p}_j + \beta_j m_e c^2
+ \sum_{j<k} V_{jk}^{ee}({\bf r}_k - {\bf r}_j) \bigg
]_{\textrm{electrons}} + \sum_{j,k} V_{jk}^{en}({\bf r}_k - {\bf
r}_j).
\end{equation}

\noindent In the first term in brackets we find a relativistic
nucleon Hamiltonian including mass, velocity, and potential terms
(including strong force interactions as well as electromagnetic
interactions).  Nuclear models of this kind can be found in the
literature \cite{Miller1972},\cite{Krutov1974}.
In the second term in brackets we find a relativistic electron
Hamiltonian also including mass, velocity and potential terms (in
this case electromagnetic interactions). Electronic models of this
kind provide the foundation for relativistic electron band
calculations; one can find a discussion of this model in \cite{Ladik1997}.
Although there is no reason to believe that a relativistic description for the
electrons is required for the problems of interest, somehow it
seems appropriate to maintain the same level of description in
the fundamental Hamiltonian.
Finally, the last term includes electron-nucleon potential
terms, which are electromagnetic here.
A further augmentation of the model to include weak interaction
physics is possible, but we will not pursue it here.

\subsection{Reduction of the nucleon Hamiltonian}

The developments presented in the previous sections allow for a
systematic reduction for the nucleon Hamiltonian in the form

$$
\bigg [ \sum_j \alphav_j \cdot c \hat{\bf p}_j + \beta_j M_j c^2 +
\sum_{j<k} V_{jk}^{nn}({\bf r}_k - {\bf r}_j) \bigg
]_{\textrm{nucleons}}
$$
\begin{equation}
~\to~ \sum_l \bigg [ {\bf M} c^2 + {|\hat{\bf P}|^2
\over 2 {\bf M}} + {\bf a} \cdot (c \hat{\bf P}) \bigg
]_l,
\end{equation}

\noindent
where the sum over nucleons is now replaced with a sum over nuclei,
and where the different matrices associated with the nuclear finite basis
expansion are selected to be appropriate for the nucleus indexed by $l$.

The excited state energies in the rest frame appear as the
eigenvalues of the mass term ${\bf M}c^2$.
The lowest-order contribution to the kinetic energy of the nucleus
as a composite Dirac particle is included in
$\displaystyle{|\hat{\bf P}|^2 / 2 {\bf M}}$, which
allows for the energy of a basis state to impact the kinetic
energy appropriately.
Finally, there is a new term ${\bf a} \cdot (c \hat{\bf P})$ that describes a relativistic effect in which
the nucleus center of mass momentum (which will subsequently be part of the
lattice vibrations) is coupled to transitions between the different basis states.
The summation over $l$ here indicates a sum over the different nuclei in the lattice, so
that there will be separate mass matrices and
lattice coupling terms for each nuclei.
The condensed matter Hamiltonian that results is


$$
\hat{H}
~=~
\left [
\sum_l
\bigg [
{\bf M} c^2
+
{|\hat{\bf P}|^2 \over 2 {\bf M}} 
+
{\bf a} \cdot (c \hat{\bf P}) 
\bigg ]_l
\right ]_{nuclei}
\ \ \ \ \ \ \ \ \ \ \ \ \ \ \ \ \ \ \ \ \ \ \ \ \ \ \ \ \ \ \ \ \ 
$$
\begin{equation}
+
\bigg [
\sum_j \alphav_j \cdot c \hat{\bf p}_j
+
\beta_j m_e c^2
+
\sum_{j<k} V_{jk}^{ee}({\bf r}_k - {\bf r}_j)
\bigg ]_{electrons}
+
\sum_{j,k} V_{jk}^{en}({\bf r}_k - {\bf r}_j)
\end{equation}

\noindent
This represents a generalization of the conventional starting place for condensed matter physics calculations.

\subsection{Born--Oppenheimer approximation}

With nucleons replaced by nuclei, the resulting model is very
similar to the standard condensed matter model, and we can
similarly make use of the Born--Oppenheimer approximation to
obtain a potential model for the nuclei

\begin{equation}
\hat{H} ~=~ \sum_l \bigg [ {\bf M} c^2 + {|\hat{\bf
P}|^2 \over 2 {\bf M}} + {\bf a} \cdot (c \hat{\bf P})
\bigg ]_l + \sum_{j<k} V_{jk}^{NN}({\bf R}_k - {\bf R}_j).
\end{equation}

\noindent
This Hamiltonian is made up of individual mass, kinetic energy, and
lattice coupling terms for each nucleus individually, and augmented
now with effective potential interactions (electromagnetic plus
electronic) between the nuclei.
This model provides a generalization of the usual lattice Hamiltonian
to include nuclear mass effects and lattice coupling with the nuclei.

\subsection{Reduction to a standard lattice Hamiltonian}

In the event that the effects associated with nuclear excitation are weak,
then the new terms in the model can be dispensed with; if we
we assume

\begin{equation}
{\bf a}_l ~\to~ 0,
\end{equation}

\noindent
then we recover a model that is essentially the standard condensed
matter lattice Hamiltonian:

\begin{equation}
\hat{H} ~=~ \sum_l \bigg [ {\bf M} c^2 + {|\hat{\bf
P}|^2 \over 2 {\bf M}} \bigg ]_l + \sum_{j<k}
V_{jk}^{NN}({\bf R}_k - {\bf R}_j).
\end{equation}

\noindent
In this approximation there is no direct coupling between the nuclear excited states
and lattice vibrations.  If the nuclei are in ground states, we could replace the
mass matrices by the mass eigenvalues, which completes the reduction:

\begin{equation}
\hat{H} ~=~ \sum_l \bigg [ M c^2 + {|\hat{\bf P}|^2
\over 2 M} \bigg ]_l + \sum_{j<k} V_{jk}^{NN}({\bf R}_k
- {\bf R}_j).
\end{equation}


\section{Discussion and Conclusions}

This study was motivated by our interest in deriving a lattice Hamiltonian
from a relativistic starting Hamiltonian in which we could study the effect
of the different configuration masses on the lattice dynamics, with the
goal of developing a systematic description of the anomalies associated
with condensed matter nuclear science.
The derivation of such a Hamiltonian from the many-particle Dirac model
in particular is in general problematic due to difficulties in the separation
of relative and center of mass degrees of freedom for the relativistic problem.
We recently obtained a weaker result that seemed to suggest a systematic way
of separating center of mass and relative contributions to the energy, which
prompted us to use it to isolate nuclear center of mass and nuclear
terms in a lattice Hamiltonian.
We obtained terms for the mass energy and kinetic energy that had been expected.
In addition we obtained an additional coupling term due to the $\alphav \cdot c {\bf P}$
terms in the Dirac model; this we had not anticipated.
The presence of this new term is not an artifact of our derivation, as we found
subsequently that it can also be obtained in a straightforward manner using
perturbation theory.

The origin of this effect is that the nuclear states of the composite nucleus
transform under a boost in the many-particle Dirac model, which implies a
mixing with other states.  In the case of constant ${\bf P}$ the eigenvalue
problem is seeking to create a version of the boosted wavefunction out of
the rest frame states.  However, with the composite momentum is dynamical (as
occurs in a lattice), the model tries to develop boosted wavefunctions for
a composite momentum that keeps changing magnitude and direction, which requires
in this picture a dynamical admixture of different rest frame states.

In the end, we obtain a Hamiltonian that describes lattice
dynamics and nuclear excitation that is derived consistently from
an underlying relativistic Hamiltonian (the many-particle Dirac
model).  This model appears to be very closely related to models
for two-level systems coupled to an oscillator that we have
investigated over the years\cite{CMNS5},\cite{CMNS7},
\cite{CMNS9},\cite{CMNS10} in connection with the excess
heat effect in the Fleischmann--Pons experiment.


\appendix
\section{Nonrelativistic limit for the Transition Operator}

Transitions in this model are described by the off-diagonal matrix element

\begin{equation}
\overline{\alphav}_{fi} \cdot c \hat{\bf P}
~=~
\bigg \langle \Phi_f \bigg |
 \sum_j {m_j \over M} \alphav_j
\cdot c   \hat{\bf P} \bigg | \Phi_i \bigg \rangle.
\end{equation}

\noindent
We are interested in developing a nonrelativistic approximation for this operator which
may be useful for understanding the coupling.

\subsection{Expansion of the wavefunction}

We assume that the solution to the relative problem can be expanded in the form

\begin{equation}
\Phi ~=~ \Phi_{+++ \cdots} + \Phi_{-++ \cdots} + \Phi_{+-+ \cdots}
+ \Phi_{++- \cdots} + \cdots,
\end{equation}

\noindent
where the first term involves large component channels for all nucleons, where the
second involves a small component channel for the first particle, and so forth.

\subsection{Channel equations}

For the first term, the relative eigenvalue problem in the rest frame results
in


$$
\bigg [
E - M c^2 - 
\sum_{j<k} V_{jk}(\xiv_k-\xiv_j)
\bigg ]
\Phi_{+++ \cdots}
~=~
\sigv_1 \cdot (c \hat{\piv}_1)
\Phi_{-++ \cdots}
+
\sigv_2 \cdot (c \hat{\piv}_2)
\Phi_{+-+ \cdots}
$$
\begin{equation}
+
\sigv_3 \cdot (c \hat{\piv}_3)
\Phi_{++- \cdots}
+
\cdots
\end{equation}

\noindent
For the second term we have


$$
\bigg [
E - M c^2 + 2 m_1 c^2 - 
\sum_{j<k} V_{jk}(\xiv_k-\xiv_j)
\bigg ]
\Phi_{-++ \cdots}
~=~
\sigv_1 \cdot (c \hat{\piv}_1)
\Phi_{+++ \cdots}
+
\sigv_2 \cdot (c \hat{\piv}_2)
\Phi_{--+ \cdots}
$$
\begin{equation}
+
\sigv_3 \cdot (c \hat{\piv}_3)
\Phi_{-+- \cdots}
+
\cdots
\end{equation}

\noindent
from which we approximate the channel wavefunction as

\begin{equation}
\Phi_{-++ \cdots}
~=~
\bigg [
E - M c^2 + 2 m_1 c^2 -
\sum_{j<k} V_{jk}(\xiv_k-\xiv_j)
\bigg ]^{-1}
(\sigv_1 \cdot c \hat{\piv}_1)
\Phi_{+++ \cdots}
\end{equation}

\subsection*{Approximate transition matrix element}

To proceed we expand the transition matrix element in terms of the
different pieces


$$
\bigg \langle \Phi_f \bigg | 
 \sum_j {m_j \over M} \alphav_j  
\cdot c   \hat{\bf P} 
\bigg | \Phi_i \bigg \rangle
~=~
{m_1 \over M}
\bigg [
\bigg \langle \Phi_f(+++ \cdots) \bigg | 
(\sigv_1 \cdot c   \hat{\bf P}) 
\bigg | \Phi_i(-++ \cdots) \bigg \rangle
$$
\begin{equation}
+
\bigg \langle \Phi_f(-++ \cdots) \bigg | 
(\sigv_1 \cdot c   \hat{\bf P}) 
\bigg | \Phi_i(+++ \cdots) \bigg \rangle
\bigg ]
+
\cdots
\end{equation}

\noindent
We expect the large component to dominate, so we keep terms with a single small
component and approximate according to

$$
\bigg \langle \Phi_f \bigg |
 \sum_j {m_j \over M} \alphav_j
\cdot c   \hat{\bf P}
\bigg | \Phi_i \bigg \rangle
~=~
\ \ \ \ \ \ \ \ \ \ \ \ \ \ \ \ \ \ \ \ \ \ \ \ \ \ \ \ \ \ \ \ \ \ \ \ \ \ \ \
\ \ \ \ \ \ \ \ \ \ \ \ \ \ \ \ \ \ \ \ \ \ \ \ \ \ \ \ \ \ \ \ \ \ \ \ \ \ \ \
\ \ \ \ \ \ \ \ \ \ \ \ \ \ \ \ \ \ \ \ \ \ \ \ \ \ \ \ \ \ \ \ \ \ \ \ \ \ \ \
$$
$$
{m_1 \over M}
\bigg [
\bigg \langle \Phi_f(+++ \cdots) \bigg |
(\sigv_1 \cdot c   \hat{\bf P})
\bigg [
E - M c^2 + 2 m_1 c^2 -
\sum_{j<k} V_{jk}(\xiv_k-\xiv_j)
\bigg ]^{-1}
(\sigv_1 \cdot c \hat{\piv}_1)
\bigg | \Phi_i(+++ \cdots) \bigg \rangle
$$
$$
+
\bigg \langle \Phi_f(+++ \cdots) \bigg |
(\sigv_1 \cdot c   \hat{\piv}_1)
\bigg [
E - M c^2 + 2 m_1 c^2 -
\sum_{j<k} V_{jk}(\xiv_k-\xiv_j)
\bigg ]^{-1}
(\sigv_1 \cdot c \hat{\bf P})
\bigg | \Phi_i(+++ \cdots) \bigg \rangle
\bigg ]
$$
\begin{equation}
+
\cdots
\end{equation}

\subsection*{Taylor series expansion}

A Taylor series expansion yields


$$
\bigg [
E - M c^2 + 2 m_1 c^2 - 
\sum_{j<k} V_{jk}(\xiv_k-\xiv_j)
\bigg ]^{-1}
~=~
{1 \over 2 m_1 c^2}
\bigg [ 
1 + {E - M c^2 \over 2 m_1 c^2}  
- {1 \over 2 m_1 c^2} \sum_{j<k} V_{jk}(\xiv_k-\xiv_j)
\bigg ]^{-1}
$$
\begin{equation}
~=~
{1 \over 2 m_1 c^2}
\bigg [
1 
- 
{E - M c^2 \over 2 m_1 c^2}  
+ 
{1 \over 2 m_1 c^2} \sum_{j<k} V_{jk}(\xiv_k-\xiv_j)
+
\cdots
\bigg ]
\end{equation}

\subsection*{Leading-order contribution}

To evaluate the leading-order term in the expansion of the transition matrix element, we
require the identity

\begin{equation}
({\bf a} \cdot \sigv) ({\bf b} \cdot \sigv) ~=~ {\bf a} \cdot {\bf
b} + i \sigv \cdot {\bf a} \times {\bf b}.
\end{equation}

\noindent
This allows us to write


\begin{samepage}

$$
(\sigv \cdot c \hat{\bf P}) 
(\sigv \cdot c \hat{\piv})
+
(\sigv \cdot c \hat{\piv})
(\sigv \cdot c \hat{\bf P}) 
~=~
2 c^2 \hat{\piv}\cdot \hat{\bf P}
+
i \sigv \cdot \bigg ( \hat{\bf P} \times \hat{\piv} + \hat{\piv} \times \hat{\bf P} \bigg )
$$
\begin{equation}
~=~
2 c^2 \hat{\piv}\cdot \hat{\bf P}
\end{equation} 

\end{samepage}

\noindent
Consequently, the leading-order term vanishes

\begin{samepage}

$$
\sum_j
{m_j \over M}
\bigg [
\bigg \langle \Phi_f(+++ \cdots) \bigg |
(\sigv_j \cdot c   \hat{\bf P})
{1 \over 2 m_jc^2}
(\sigv_j \cdot c \hat{\piv}_j)
\bigg | \Phi_i(+++ \cdots) \bigg \rangle
$$
$$
+
\bigg \langle \Phi_f(+++ \cdots) \bigg |
(\sigv_j \cdot c   \hat{\piv}_j)
{1 \over 2 m_jc^2}
(\sigv_j \cdot c \hat{\bf P})
\bigg | \Phi_i(+++ \cdots) \bigg \rangle
\bigg ]
$$
\begin{equation}
~=~ {1 \over M} \bigg [ \bigg \langle \Phi_f(+++ \cdots) \bigg |
\hat{\bf P} \cdot \sum_j \hat{\piv}_j \bigg | \Phi_i(+++ \cdots)
\bigg \rangle,
\end{equation},

\end{samepage}

\noindent
since

\begin{equation}
\sum_j\piv_j ~=~ \sum_j {\bf p}_j - {m_j \over M} {\bf P} ~=~ 0.
\end{equation}

\subsection*{First next-order term}

There are two next-order terms; the first of these is

$$
\sum_j
2
{m_j \over M}
\bigg \langle \Phi_f(+++ \cdots) \bigg |
(\sigv_j \cdot c   \hat{\bf P})
{E - M c^2 \over (2 m_j c^2)^2}
(\sigv_j \cdot c \hat{\piv}_j)
\bigg | \Phi_i(+++ \cdots) \bigg \rangle
\ \ \ \ \ \ \ \ \ \ \ \ \ \ \ \ \ \ \ \
$$
\begin{equation}
~=~ {(E - M c^2) \over 2 Mc^2} \bigg \langle \Phi_f(+++ \cdots)
\bigg | \sum_j { \hat{\piv_j} \cdot \hat{\bf P} \over m_j } \bigg
| \Phi_i(+++ \cdots) \bigg \rangle.
\end{equation}

\noindent
Note that this term contains the sum $\sum_j { \hat{\piv_j} \over m_j }$
which we may rewrite as

\begin{equation}
\sum_j { \hat{\piv_j} \over m_j } ~=~ \sum_j \hat{\piv}_j \left (
{1 \over m_j} - {1 \over m_{av}} \right ),
\end{equation}

\noindent
since $\sum_j \hat{\piv}_j = 0$, where $m_{av}$ might be an appropriate average of the
different masses.  Since the proton mass is 938.27 MeV/c$^2$ and the
neutron mass is 939.56 MeV/c$^2$, the individual nucleon masses are little different
from the average nucleon mass.  Consequently, we expect this term to be small.

\subsection*{Second next-order term}

The other next-order term is

$$
\sum_j
{m_j \over M}
\bigg [
\bigg \langle \Phi_f(+++ \cdots) \bigg |
(\sigv_j \cdot c   \hat{\bf P})
\bigg [
{1 \over (2 m_j c^2)^2}
\sum_{k<l} V_{kl}(\xiv_l-\xiv_k)
\bigg ]
(\sigv_j \cdot c \hat{\piv}_j)
\bigg | \Phi_i(+++ \cdots) \bigg \rangle
$$
$$
+
\bigg \langle \Phi_f(+++ \cdots) \bigg |
(\sigv_j \cdot c   \hat{\piv}_j)
\bigg [
{1 \over (2 m_j c^2)^2}
\sum_{k<l} V_{kl}(\xiv_l-\xiv_k)
\bigg ]
(\sigv_j \cdot c \hat{\bf P})
\bigg | \Phi_i(+++ \cdots) \bigg \rangle
\bigg ]
$$
$$
~=~
{1 \over 2 Mc^2}
\bigg [
\bigg \langle \Phi_f(+++ \cdots) \bigg |
\sum_j
(\sigv_j \cdot c   \hat{\bf P})
\bigg [
{1 \over 2 m_j c^2}
\sum_{k<l} V_{kl}(\xiv_l-\xiv_k)
\bigg ]
(\sigv_j \cdot c \hat{\piv}_j)
\bigg | \Phi_i(+++ \cdots) \bigg \rangle
$$
\begin{equation}
+ \bigg \langle \Phi_f(+++ \cdots) \bigg | \sum_j (\sigv_j \cdot c
\hat{\piv}_j) \bigg [ {1 \over 2 m_j c^2} \sum_{k<l}
V_{kl}(\xiv_l-\xiv_k) \bigg ] (\sigv_j \cdot c \hat{\bf P}) \bigg
| \Phi_i(+++ \cdots) \bigg \rangle \bigg ].
\end{equation}

\end{document}